\documentclass[runningheads]{llncs}
\usepackage[T1]{fontenc}
\usepackage{makecell}
\usepackage{inputenc}
\usepackage{subcaption}
\usepackage{tikz}
\usepackage{hyperref}       
\usepackage{url}            
\usepackage{booktabs}       
\usepackage{amsfonts}       
\usepackage{nicefrac}       
\usepackage{microtype}      
\usepackage{lipsum}
\usepackage{graphicx}
\usepackage{listings}
\usepackage{fontspec}
 \usepackage{amsmath} 
 \usepackage{longtable,array}
\usetikzlibrary{mindmap,trees,shadows}
\usetikzlibrary{positioning}
\usetikzlibrary{arrows.meta,positioning,calc,decorations.pathreplacing}
\usepackage[compatibility=false]{caption}
\usepackage{tcolorbox}
\usepackage{xcolor}

\begin{document}
\newtcolorbox{rqanswer}{
  colback=gray!5,
  colframe=black!60,
  boxrule=0.5pt,
  arc=2pt,
  left=6pt,
  right=6pt,
  top=6pt,
  bottom=6pt
}
\definecolor{CodeBackGround}{RGB}{240, 240, 240} 

\setmonofont[
  Path=fonts/,
  UprightFont=APL385.ttf
]{APL385}

\lstdefinelanguage{apl}{
  morekeywords={⍺⍵⍶⍹←→∆∇⍺⍵},
  sensitive=true
}
\definecolor{CodeComment}{RGB}{0,128,0}
\definecolor{CodeBackGround}{RGB}{245,245,245}

\makeatletter
\lst@InputCatcodes
\def\lst@DefEC{%
 \lst@CCECUse \lst@ProcessLetter
  ^^80^^81^^82^^83^^84^^85^^86^^87^^88^^89^^8a^^8b^^8c^^8d^^8e^^8f%
  ^^90^^91^^92^^93^^94^^95^^96^^97^^98^^99^^9a^^9b^^9c^^9d^^9e^^9f%
  ^^a0^^a1^^a2^^a3^^a4^^a5^^a6^^a7^^a8^^a9^^aa^^ab^^ac^^ad^^ae^^af%
  ^^b0^^b1^^b2^^b3^^b4^^b5^^b6^^b7^^b8^^b9^^ba^^bb^^bc^^bd^^be^^bf%
  ^^c0^^c1^^c2^^c3^^c4^^c5^^c6^^c7^^c8^^c9^^ca^^cb^^cc^^cd^^ce^^cf%
  ^^d0^^d1^^d2^^d3^^d4^^d5^^d6^^d7^^d8^^d9^^da^^db^^dc^^dd^^de^^df%
  ^^e0^^e1^^e2^^e3^^e4^^e5^^e6^^e7^^e8^^e9^^ea^^eb^^ec^^ed^^ee^^ef%
  ^^f0^^f1^^f2^^f3^^f4^^f5^^f6^^f7^^f8^^f9^^fa^^fb^^fc^^fd^^fe^^ff%
  ^^^^20ac^^^^0153^^^^0152%
  ^^^^20a7^^^^2190^^^^2191^^^^2192^^^^2193^^^^2206^^^^2207^^^^220a%
  ^^^^2218^^^^2228^^^^2229^^^^222a^^^^2235^^^^223c^^^^2260^^^^2261%
  ^^^^2262^^^^2264^^^^2265^^^^2282^^^^2283^^^^2296^^^^22a2^^^^22a3%
  ^^^^22a4^^^^22a5^^^^22c4^^^^2308^^^^230a^^^^2336^^^^2337^^^^2339%
  ^^^^233b^^^^233d^^^^233f^^^^2340^^^^2342^^^^2347^^^^2348^^^^2349%
  ^^^^234b^^^^234e^^^^2350^^^^2352^^^^2355^^^^2357^^^^2359^^^^235d%
  ^^^^235e^^^^235f^^^^2361^^^^2362^^^^2363^^^^2364^^^^2365^^^^2368%
  ^^^^236a^^^^236b^^^^236c^^^^2371^^^^2372^^^^2373^^^^2374^^^^2375%
  ^^^^2377^^^^2378^^^^237a^^^^2395^^^^25af^^^^25ca^^^^25cb%
  ^^00}
\lst@RestoreCatcodes
\makeatother

\lstset{%
  basicstyle=\ttfamily\small,  
  identifierstyle=,                             
  commentstyle=\slshape\color{CodeComment},    
  stringstyle=\ttfamily,                        
  showstringspaces=false,                      
  backgroundcolor=\color{CodeBackGround},      
  frame=single,                               
  framesep=1pt,                                
  framerule=0.8pt,                              
  rulecolor=\color{CodeBackGround},             
  breaklines=true,                            
  breakindent=0pt,
  inputencoding=utf8,
  extendedchars=true
}
\lstdefinestyle{pythonstyle}{
  language=Python,
  basicstyle=\ttfamily\small,
  keywordstyle=\color{blue},
  stringstyle=\color{orange},
  commentstyle=\color{green!50!black},
  numbers=left,
  numberstyle=\tiny,
  stepnumber=1,
  numbersep=8pt,
  showstringspaces=false,
  breaklines=true,
  frame=single,
  tabsize=2
}

\title{Neural Code Translation of Legacy Code: \\ APL to C\#}

\author{Abdulrahman Ramadan\inst{1} \and
Hanen Borchani \inst{2} \and
Iben Lilholm\inst{2} \and 
Mikkel Almind\inst{2} \and 
Allan Peter Engsig-Karup \inst{1,2}}
\authorrunning{A. Ramadan et al.}
%
\institute{Department of Applied Mathematics and Computer Science,\\ Technical University of Denmark, Denmark \\
\email{abdulrahman.ramdan@outlook.com}
\and
SimCorp A/S, Copenhagen, Denmark}
\maketitle   

\begin{abstract}
Automatic translation between programming languages remains a challenging problem, particularly when the source language is highly concise and specialized. This paper investigates the translation of APL into C\# using large language models. The task is difficult due to APL’s sparse syntax, the scarcity of large-scale parallel corpora, and the requirement for specialized knowledge to interpret APL programs. To address these challenges, we introduce a novel framework for APL-to-C\# translation by comparing three guided strategies, namely natural language description-mediated, retrieval-augmented, and iterative refinement, against a baseline direct translation model. We constructed multiple datasets of functionally equivalent code pairs spanning various levels of complexity, and to rigorously assess translation quality, we developed an automated evaluation pipeline that verifies both syntactic compilation and functional execution of the generated C\# code. Our results demonstrate that neural code translation can successfully bridge the gap between APL and C\# for a wide range of programs, and that incorporating additional context and guidance significantly improves model performance.
\end{abstract}


\section{Introduction}

Many organizations rely on legacy systems implemented in specialized and less commonly used programming languages, which poses significant challenges for long-term maintainability and integration with modern software architectures. Furthermore, as the pool of experts familiar with these languages diminishes, there is a growing and ongoing need for reliable automated methods to facilitate the migration of legacy code to modern programming languages. In this study, we take the view that legacy code refers to old or inherited software code that is still in use but is difficult to maintain, update, or extend. With a view to the recent decades progress in natural language processing technology, we seek to investigate modern means to address such high-value tasks.

During the past decade, the rise of large language models (LLMs) based on the transformer architecture \cite{Vaswani2023}, originally designed for human language translation tasks, has shown promising results in a variety of software engineering tasks \cite{Wu2016}. However, recent research has shown that while certain problems can be reliably solved by such LLMs, others, though seemingly similar in complexity, remain beyond the current capabilities of these models \cite{DellAcqua2023}. 

This study evaluates the efficacy of LLMs in automating the translation of legacy APL \cite{Iverson1962} to C\#, establishing a benchmark for their current capabilities in this domain. To address the scarcity of large-scale parallel corpora, we follow an approach similar to the recent high-performance computing translation of OpenMP Fortran to C\# \cite{Lei2023} by constructing dedicated datasets of functionally equivalent APL and C\# program pairs that span diverse levels of abstraction and complexity in APL code. Leveraging these datasets, fine-tuning pre-trained LLMs using parameter efficient techniques such as LoRA \cite{Hu2022} can help enhance performance on specialized translation tasks. Furthermore, other techniques that involve instruction-tuned models, used in iterative loops for automatic program repair \cite{Ruiz2025} offer a viable path for reducing the manual effort required to debug and refine the resulting source code. 
Lastly, large language models can introduce errors when translating code between programming languages~\cite{pan2023lost}, we therefore further investigate the dominant types of translation errors produced during this process.
Consequently, this work is structured around the following research questions regarding the neural code translation of APL to C\#:

\begin{description} 
\item[RQ1: Translation quality and guidance strategies:] How do various LLMs perform in functional APL-to-C\# translation, and which guided approaches most effectively enhance their performance?

\item[RQ2: Training data scaling effect:] To what extent is the model performance limited by the scale of the training data?

\item[RQ3: Translation error analysis:] What are the primary root causes of translation failure when translating APL to C\#?
\end{description}

The remainder of this paper is organized as follows. Section~\ref{sec:related_work} reviews the existing related works, and Section~\ref{sec:background} briefly presents APL foundations and the main challenges inherent in APL-to-C\# translation. Section~\ref{sec:methodology} introduces our methodology, while Section~\ref{sec:experiment-setup} covers our experimental setup. Section~\ref{sec:experiment-results} discusses the experimental results and overall LLMs performance. Finally, Section~\ref{sec:conclusion} concludes the article and suggests directions for future work.

\section{Related Work}
\label{sec:related_work}

The task of automatic source code translation has been studied in recent years, as there is a demand for legacy code migration. The quality of code translation depends on the choice of source and target programming languages \cite{Aljagthami2025} meaning that findings from one language pair may not generalize to others.

A review of the literature \cite{Chen2025} shows that most code translation studies and benchmarks focus on commonly used programming languages. For instance, most datasets benchmarks such as CodeXGLUE \cite{Lu2021}, CodeNet \cite{Puri2021}, Avatar \cite{Ahmad2023Avatar}, and Chen et al. \cite{Chen2018}, are used to benchmark code translation tasks across common programming languages. Specifically, CodeXGLUE includes Java-C\# pairs, CodeNet emphasizes C++, Python, and Java from various programming competitions, Avatar focuses on Java-Python pairs, and Chen et al. address Java–C\# translation. Consequently, less commonly used languages such as APL remain largely absent from the literature. Industrial code migration tools for niche languages are also emerging, such as IBM's watsonx Code Assistant for Z \cite{IBM2023}, which is designed to support the transformation of legacy COBOL code into Java. 

In addition, researchers have introduced several approaches to enhance translation quality where data is limited. One prominent approach utilizes natural language as an intermediate representation; for instance, Ahmad et al. \cite{Ahmad2023Backtranslate} leverages code summarization and generation for Java-to-Python and Python-to-Java translations, while Tai et al. \cite{Tai2025NL} shows that combining chain-of-thought prompting with intermediate natural language summaries yields superior translation performance. Their study evaluated translation pairs across C++, C, Python, Java, and Go languages. Retrieval-Augmented Generation (RAG) has also been used by Bhattarai et al. \cite{Bhattarai2024} with few-shot learning to guide translation from Fortran to C++ by dynamically retrieving the most relevant examples from a repository of existing code translations and providing them to the model context.

Beyond single-pass generation, iterative translation frameworks have been proposed to ensure functional equivalence between the source and target code. Tools such as AlphaTrans \cite{Ibrahimzada2025} implement iterative feedback-based prompting loops, incorporating error descriptions whenever  generated code validation fails. This approach was specifically demonstrated for Java-to-Python translation.

In general, existing literature focuses primarily on commonly used programming languages, leading to a lack of specialized datasets and evaluation frameworks for array-oriented languages such as APL. To the best of our knowledge, no existing work has investigated the neural code translation of APL into a modern language. In this study, we address this gap by presenting the first systematic investigation and specialized dataset for APL-to-C\# translation, providing a foundational resource for the modernization of legacy APL systems. To overcome the challenge of limited data availability, we validate our approach by translating code from a real-world production system, reporting on practical metrics that highlight the efficacy and applicability of our methodology.

\section{Background: APL Foundations and APL-to-C\# Translation Challenges}
\label{sec:background}

\subsection{APL Language Overview}
\label{sec:overview_of_apl}
APL is an array-oriented language developed by Kenneth E. Iverson \cite{Iverson1962} as a notation for algorithmic reasoning without explicit loops or indexing. Its core data type is the multidimensional array, with most primitives defined via array operations. In this study, we use Dyalog APL \cite{dyalog_oo_apl}, a widely used implementation that extends the language with object-oriented constructs for large-scale systems. 

APL primitives operate directly on multidimensional arrays without explicit loops. For example, \texttt{2 3 ⍴ ⍳6} generates integers from 1 to 6 using \texttt{⍳} primitive and reshapes them into a $2 \times 3$ matrix using \texttt{⍴} primitive. This matrix can be transposed using \texttt{⍉} primitive: 
\begin{lstlisting}[basicstyle=\ttfamily\scriptsize]
    ⍉ 2 3 ⍴ ⍳6
1 4
2 5
3 6
\end{lstlisting}
\vspace{-5pt}

Reductions are performed by applying the \texttt{/} operator to another operator. For example, \texttt{×/V} and \texttt{⌈/V} compute the product and maximum of the vector $V$, respectively.

\setlength{\multicolsep}{0pt}
\begin{multicols}{2}
\begin{lstlisting}[basicstyle=\ttfamily\scriptsize]
    ×/ 3 7 2 5
210
\end{lstlisting}
\columnbreak
\begin{lstlisting}[basicstyle=\ttfamily\scriptsize]
    ⌈/ 3 7 2 5
7
\end{lstlisting}
\end{multicols}

APL functions are \textit{monadic} (taking a single right argument, $\omega$) or \textit{dyadic} (taking both a left argument, $\alpha$, and a right argument, $\omega$). Examples include the following monadic $Rank$ and dyadic $IndexSelect$ functions:
\noindent
\begin{multicols}{2}
\begin{lstlisting}[basicstyle=\ttfamily\scriptsize]
    Rank ← { ⍴⍴⍵ }
    Rank 2 3 4 ⍴⍳24
3
\end{lstlisting}
\columnbreak
\begin{lstlisting}[basicstyle=\ttfamily\scriptsize]
    IndexSelect ← { ⍺⊃ ¨ ⊂⍵ }
    1 3 5 IndexSelect 'ABCDE'
ACE
\end{lstlisting}
\end{multicols}
Standard APL constructs are often combined into recognizable entities known as \textit{APL idioms}. For example, the average of an array \texttt{X} is expressed as \texttt{(+⌿X)÷≢X}. Finally, to manage APL's dynamic typing in large-scale systems, developers use \textit{function APL headers} that document expected argument shapes and types. For instance, the following APL header presents a dyadic function that operates on integer arrays and returns an integer: 

\begin{lstlisting}[basicstyle=\ttfamily\scriptsize]
 ⍝ ⍺ : INT[]   ⍵ : INT[]   → INT
\end{lstlisting}

\subsection{Challenges in APL-to-C\# Translation}
\label{sec:translation_challenges}

Translating APL to C\# requires bridging two fundamentally different paradigms. As a \textit{statically typed}, \textit{object-oriented language}, C\# is structured around classes and procedural logic designed for maintainability in large-scale systems. In contrast, APL is a \textit{dynamically typed}, \textit{array-oriented language} that treats computation as mathematical transformations on entire data structures rather than sequences of operations on individual elements. This shift introduces significant structural and type-mapping challenges.

APL handles scalars, vectors, character arrays, and matrices uniformly, allowing the same function to run across various dimensionalities without the need of explicit loops or index manipulation. This introduces significant ambiguity when translating to C\#, where types are distinct and functions are typically bound to specific signatures. Consequently, mapping APL to C\# requires function overloading and extensive runtime type-checking, which complicates the implementation and can obscure the original logic. To illustrate this, consider the following APL function which performs membership checks between values and collections of values of various dimensionalities. It has two arguments that can be of any type or rank:

\begin{lstlisting}[basicstyle=\ttfamily\scriptsize]
r ← y xMsInt x
 r ← y ∊ x
\end{lstlisting}

Even when limiting the scope to integers with a maximum rank of 2, a C\# translation requires method overloading for every possible combination of argument types. This means that the translation should include an implementation of each of the following method signatures: 

\begin{lstlisting}[basicstyle=\ttfamily\scriptsize]
    public bool xMsInt(int y, int[] x)
    public bool[] xMsInt(int[] y, int x)
    public bool[] xMsInt(int[] y, int[] x)
    public bool[,] xMsInt(int[,] y, int x)
    public bool[,] xMsInt(int[,] y, int[] x)
    public bool xMsInt(int y, int[,] x)
    public bool[] xMsInt(int[] y, int[,] x)
    public bool[,] xMsInt(int[,] y, int[,] x)
\end{lstlisting}

Alternatively, static type distinctions via method overloading could be replaced with explicit runtime checks. Instead of relying on the C\# compiler to select an overload based on declared argument types, a single method can be defined to accept generic \texttt{object} parameters. In this case, types must be inspected at execution time. A sketch of a method using this approach is as follows:

\begin{lstlisting}[basicstyle=\ttfamily\scriptsize]
public static object xMsInt(object y, object x)
{
    switch (y, x)
    {
        case (int ys, int[] xa): ...       // returns bool
        case (int[] ya, int xs): ...       // returns bool[]
        case (int[] ya, int[] xa): ...     // returns bool[]
        case (int[,] ym, int xs): ...      // returns bool[,]
        case (int[,] ym, int[] xa): ...    // returns bool[,]
        case (int ys, int[,] xm): ...      // returns bool
        case (int[] ya, int[,] xm): ...    // returns bool[]
        case (int[,] ym, int[,] xm): ...   // returns bool[,]
    }
}
\end{lstlisting}

As shown, translating APL to C\# often results in significantly more verbose implementations.  Consequently, a single line of APL expressing an array transformation often expands into a complex C\# code involving loops, bounds checks, and intermediate variables.  Another distinction is the use of indexing: APL uses 1-based indexing, whereas C\# uses 0-based indexing. This can introduce off-by-one errors, when indices are returned or passed as arguments, and compromise functional equivalence. These differences make direct translation challenging and prone to producing code that obscures the original intent.

\section{Methodology}
\label{sec:methodology}

In this section, we present the methodology used for the translation of APL-to-C\#. We treat code translation as a {\em sequence-to-sequence generation problem} and use LLMs to perform the translation. To address the structural and type ambiguities described in Section~\ref{sec:translation_challenges}, we investigate the use of translation guiding methods to enhance the generated C\# translations.
As for evaluation, we use a compile-and-execute setup, where the correctness is measured based on whether the generated code compiles and produces the expected outputs on test cases.

\subsection{Translation Approaches}
\label{sec:trans_appraoches}

We consider the following four translation approaches: direct translation, translation with natural language description, retrieval-augmented translation, and iterative translation.
Note that we do not consider training a language model from scratch since this is generally too costly and requires a significant sized dataset of functionally equivalent APL-C\# pairs. Instead, we focus on fine-tuning and prompting strategies.

\subsubsection{Direct Translation (Baseline)}
In this approach, the model is fine-tuned to translate APL code directly to its C\# equivalent. Each training instance is a pair of an APL program with its functionally corresponding C\# implementation. At inference time, the model is provided an APL code snippet and is tasked with generating the C\# translation.

\subsubsection{Translation with Natural Language Description}
Previous research by Tai et al. \cite{Tai2025NL} shows that incorporating natural language (NL) descriptions can improve code translation quality. Based on this, we investigate this approach using two main steps: first, an LLM is used to generate a high-level natural description of the APL source code; then, a second LLM uses this description and the APL code to generate the C\# translation. This approach can be executed as either a single generation task or as two distinct stages. We evaluate several training and inference configurations to compare the effectiveness of pre-trained (PT) off-the-shelf models against fine-tuned (FT) ones. Specifically, we explore the following configurations:

\begin{itemize}
\item APL $\xrightarrow{\text{FT}}$ NL, APL+NL $\xrightarrow{\text{FT}}$ C\#: Both stages are fine-tuned. The first model is trained on APL-to-NL pairs, and the second on (APL+NL)-to-C\# pairs.

\item APL $\xrightarrow{\text{FT}}$ NL, APL+NL $\xrightarrow{\text{PT}}$
C\#: Only the first stage is fine-tuned where the model is trained on APL-to-NL pairs. The second stage uses a pre-trained model to generate C\# code from the combined APL and NL inputs.

\item APL $\xrightarrow{\text{PT}}$ NL, APL+NL $\xrightarrow{\text{FT}}$ C\#: The first stage uses a pre-trained model to generate the NL description, which is then passed to a second model fine-tuned on (APL+NL)-to-C\# pairs.

\item APL $\xrightarrow{\text{PT}}$ NL, APL+NL $\xrightarrow{\text{PT}}$ C\#: Both stages use pre-trained LLMs; the first generates the NL description, and the second generates C\# code from the combined APL and NL inputs.
\end{itemize}

\subsubsection{Retrieval-Augmented Translation} 
We also investigated the use of Retrieval-Augmented Generation (RAG) \cite{Lewis2021} by incorporating a corpus of external APL documentation to mitigate ambiguity during the translation process. For each input APL code snippet, relevant documentation is retrieved via vector similarity search and provided to the LLM as additional context.

\subsubsection{Iterative Translation}
Finally, we consider an iterative translation strategy where compilation and execution feedback are used to refine the translation results. In fact, after the initial C\# code is produced, it is compiled and executed against a predefined set of test cases. Any resulting compilation or runtime errors are fed back to the model as additional context for a subsequent refinement step. This process is repeated for a fixed number of iterations or until a correct functional C\# code is generated. 

\subsection {Dataset Construction}
The dataset used for supervised fine-tuning and evaluation consists of APL programs aligned with functionally equivalent C\# implementations. For the NL-based approaches, each data point additionally includes an LLM-generated description of the APL program. Listing~\ref{lst:datapoint_example} illustrates a representative data point. Note that $apl$, $csharp$, and $nl\_description$ fields are used during the training phase, while $io~(input-output)$, including test cases with paired inputs and expected outputs, is only used for validation.

\begin{lstlisting}[caption={Example of a datapoint.},label={lst:datapoint_example},basicstyle=\ttfamily\scriptsize,]
  {
    "apl": " r←y xIsectr x ...",
    "csharp": "public class xIsectrUtil ...",
    "io": [
      {
        "method_name": "xIsectr",
        "AplLeftArg": "2 2 ⍴ 3 4 1 2",
        "AplRightArg": "2 2 ⍴ 1 2 3 4",
        "CSharpArg": "new object[,] { { 3, 4 }, { 1, 2 } }, new   object[,] { { 1, 2 }, { 3, 4 } }",
        "Output": [[1, 2], [3, 4]]
      },
      {
        "method_name": "xIsectr",
        "AplLeftArg": "2 2 ⍴ 3 5 1 3",
        "AplRightArg": "2 2 ⍴ 1 2 3 4",
        "CSharpArg": "new object[,] { { 3, 5 }, { 1, 3 } }, new object[,] { { 1, 2 }, { 3, 4 } }",
        "Output": []
      }
    ],
    "nl_description": "This function computes the intersection of two matrices, **y** and **x**. \nIt takes the following inputs ..."
  }
\end{lstlisting}

The datasets were constructed from various sources, with varying complexity. 

\begin{itemize}
    \item{\textbf{Dataset $A$: Initial Curated Dataset.}} Dataset~$A$ consists of $|A|=800$ manually curated APL--C\# pairs covering fundamental language constructs.

    \item{\textbf{Dataset $B$: Utility APL Functions.}} Dataset~$B$ includes $|B|=143$ manually curated utility functions derived from production environments. In contrast to Dataset $A$, this dataset contains more complex examples.

    \item{\textbf{Dataset $C$: Cross-Language Public Implementations.}}  Dataset~C consists of $|C| = 320$ implementations sourced from Rosetta Code \cite{rosetta_code}. For each task, an APL solution and a corresponding C\# solution were selected.

    \item{\textbf{Dataset $I$: APL Idioms.}} Dataset~I contains $|I| = 45$ APL idioms. Each idiom is associated with a description explaining its intention. The dataset was constructed using these descriptions to generate the corresponding implementations of C\#. The resulting code was then manually verified.
\end{itemize}

Each dataset was partitioned into training, validation, and test splits. For model fine-tuning, we aggregated the training subsets across all datasets, resulting in a total training corpus of $1066$ samples. We observed that the contribution of each dataset to model performance was non-uniform. Specifically, Dataset $B$, which comprises production-grade code, had the strongest influence on the accuracy of APL-to-C\# translations. Consequently, we utilized the test split of Dataset $B$ ($|B_{test}| = 49$) to evaluate the generalization capabilities of the model in production-oriented tasks.

In addition to the curated datasets, we consider the use of \textbf{method signatures} in the prompts. In fact, code generation using LLMs alone often fails to guaranty the strict typing requirements of C\#, which requires precise parameter and return types for successful compilation and execution. To address this, we developed an APL header parser using the functional programming language F\#. This tool extracts type information from APL function headers and generates corresponding C\# method signatures, which are then included in the prompt to guide the model’s generation process.

\section{Experimental Setup} 
\label{sec:experiment-setup}
\subsection{Analysis of Models Tokenization}
\label{sec:model-tokenization-analysis}
Since APL glyphs are not very common in most LLM training datasets, we analyzed how different tokenizers handle APL glyphs. Specifically, we evaluated the efficiency of different tokenizers selected from Hugging Face \cite{huggingface}. Table~\ref{tab:apl_tokenization_stats} shows the tokenization performance of these models based on the single-token rate, the average tokens per glyph and per sample, and round-trip failures, defined as instances where encoding and decoding fails to reproduce the original input.

\begin{table}[th!]
\centering
\caption{Tokenization efficiency and reconstruction integrity across models.}
\label{tab:apl_tokenization_stats}
\begin{tabular}{lcccc}\hline
Model
& \makecell{APL Single\\Token Rate}
& \makecell{Avg Tokens\\ per Glyph}
& \makecell{Avg Tokens\\ per Sample}
& \makecell{Round-trip\\ Failures} \\
\hline
Qwen3-32B & 0.715 & 1.284 & 262.274 & 0 \\
Gemma-4-31b-it & 0.671 & 1.656 & 277.475 & 1 \\
Deepseek-Coder-6.7b-Instruct & 0.366 &  2.361 & 315.243 & 61\\
CodeLlama-34b-hf & 0.350 &  2.352 & 296.834 & 1\\
\hline
\end{tabular}
\end{table}

We found that Gemma4-30B-it\cite{gemma4} and Qwen3-32B \cite{yang2025qwen3technicalreport} perform the best and have the most efficient tokenization, with 70\% of APL glyphs represented as single tokens and the lowest average token count per sample. In contrast, other models require significantly more tokens per glyph, resulting in higher overall sequence lengths. Furthermore, Deepseek-Coder-6.7b-instruct tokenizer fails to preserve the division symbol \verb|÷| replacing it with the unknown symbol \verb|<unk>|, resulting in a high number of round-trip failures. The single failure reported for Gemma-4-31b-it and CodeLlama-34b-hf are due to unrelated symbols in the dataset rather than APL tokenization error. This analysis shows that robust handling of APL glyphs is an important performance factor as LLMs with better APL tokenization produce shorter sequences and avoid round-trip failures.

\subsection{Model Selection and Setup}
In addition to the open-weight models, \textbf{Gemma4-30B-it} and \textbf{Qwen3-32B} (selected based on the tokenization analysis in Section~\ref{sec:model-tokenization-analysis}), we evaluate several closed-weight models, namely \textbf{GPT-5} from OpenAI \cite{gpt5_chat} and \textbf{Claude Opus 4} \cite{anthropic_claude4} from Anthropic. We consider the four following setups:

\begin{itemize}
\item{\textbf{Direct APL-to-C\# Translation.}} In this baseline setup, the model is fine-tuned to generate C\# code directly from an input APL program without any guidance or intermediate steps.

\item{\textbf{APL-to-C\# Translation with Natural Language Description.}} In this setup, we use a sequential pipeline using two instances of the open-weight model. The first instance was fine-tuned to generate NL descriptions from the APL code, while the second was fine-tuned to generate C\# code from a combination of these generated NL descriptions and the APL code.

\item{\textbf{Retrieval-Augmented APL-to-C\# Translation.}} Here, we utilize a curated set of documents containing APL idioms and their corresponding explanations, collected from both the APL2 Idioms Library \cite{apl2_idioms} and the FinnAPL Idiom Library \cite{finnapl}. Each document is partitioned into text chunks and vectorized using the text-embedding-3-large model \cite{openai_embedding}. At inference time, the source APL snippet serves as the retrieval query, and the retrieval engine computes cosine similarity scores to identify and retrieve the top 5 most relevant context chunks. To manage the context window efficiently, these relevant chunks are first condensed into a concise summary using the lightweight GPT-5-mini model \cite{gpt5_mini}, and this summary is then provided as additional context to GPT-5 \cite{gpt5_chat} to produce the final C\# translation.

\item{\textbf{Iterative APL-to-C\# Translation.}} This setup is implemented as a wrapper around the translation model and the compile-and-execute environment. In each iteration, the generated C\# code is validated against the provided input--output test cases. If any test fails, the system initiates a subsequent iteration by augmenting the prompt with a historical record of previous attempts. This includes prior code snippets as well as per-test results, specifically: compiler error messages, method arguments and signature, expected and actual outputs. This iterative process continues for up to 5 iterations or until a correct translation output is achieved.

\end{itemize}

\paragraph{\textbf{Fine-tuning and Inference Setup.}} The models were fine-tuned using an instruction prompt that defines the translation task and provides formatting constraints for the output. Fine-tuning requires a model-specific chat template to define how different roles are arranged within the input sequence; an example of this structure is illustrated in Listing.~\ref{lst:prompt_roles}. To ensure experimental consistency, the same prompt content was maintained in all models. During inference, this prompt was extended with signatures of the C\# method. In the translation with NL description setup, the program’s NL description was additionaly included.

\begin{lstlisting}[caption={Example of an instruction prompt used for fine-tuning.},label={lst:prompt_roles},basicstyle=\ttfamily\scriptsize, breaklines=true]
<|im_start|>system
You are an expert APL code programmer.
Given the following APL code create C# program that implements the given code. Output only the C# program, with no example usage.<|im_end|>

<|im_start|>user
### APL code:
r←or v
r←0
:For e :In v
    r∨←e
    :If r=1 ⋄ :Leave ⋄ :EndIf
:EndFor
<|im_end|>

<|im_start|>assistant
Output format: Only compilable C# program code, no explanations, no reasoning, no example usage.
### C#:
\end{lstlisting}

\subsection{Practical Limitations}
All experiments in this study were conducted on high-performance computing systems from the Technical University of Denmark. The used node was equipped with 2 Nvidia H100 PCIe GPUs with 80 GB VRAM each. We configured batch jobs to restrict the allocation of 2 GPUs per job. Consequently, we were restricted to only selecting models that can be accommodated within these compute and memory thresholds. 

The limitations of VRAM are particularly significant during fine-tuning, as memory must accommodate model parameters, intermediate activations, optimizer states, and gradients simultaneously. To further mitigate the memory footprint, we used Parameter-Efficient Fine-Tuning methods, specifically LoRA \cite{Hu2022} and 8-bit quantization. Although quantization effectively reduced the memory required for model weights, the associated dequantization overhead increased fine-tuning durations several times compared to standard training. 

Furthermore, an analysis of the tokenized training data revealed that many samples exceeded 1024 tokens; therefore, we used a sequence length of 2048 for fine-tuning and inference. Since the memory requirements of the attention mechanism scale quadratically with the sequence length, this required a higher VRAM allocation per sample. Given these constraints, the largest model viable for fine-tuning in our setup had approximately 30 billion parameters. Fine-tuning of the complete training set required typically 9–12 hours, while inference across the entire test dataset was completed in roughly 15 minutes.

\section{Experimental Results}
\label{sec:experiment-results}

\subsection{Overall Translation Performance} 

Table~\ref{tab:compilation_pass_rates} presents the compilation and execution performance in all evaluated translation setups. The results indicate that introducing signatures significantly improves overall performance, substantially increasing the models’ full pass rates. Fine-tuned Gemma-4-31b-it emerges as a strong performer, rivaling closed-weights alternatives. In addition, the inclusion of NL descriptions consistently improves both partial- and full pass rates across all models. This shows that providing explicit type and shape information is important for handling APL translation.

\begin{table}[th!]
\centering
\caption{Compilation and execution pass rates across translation setups. \textit{Compile rate} denotes the fraction of programs that compile successfully; \textit{Full pass rate} indicates those passing all test cases; \textit{Partial pass rate} indicates those passing a subset. \textit{FT}: Fine-tuned; \textit{Signatures}: Use of APL header parser (type/shape info).}
\label{tab:compilation_pass_rates}
\resizebox{\textwidth}{!}{%
\begin{tabular}{lccc}
\toprule
\textbf{Setup/Model} 
& \textbf{Compile Rate} 
& \textbf{Partial Pass Rate} 
& \textbf{Full Pass Rate} \\
\midrule
\multicolumn{4}{l}{\textbf{\textit{Direct APL-to-C\# Translation}}} \\
\midrule

Qwen3-32B (\textbf{FT})
& 81.63\% (40/49)
& 38.78\% (19/49)
& 18.37\% (9/49) \\

Qwen3-32B (\textbf{FT}, \textbf{Signatures})
& 95.92\% (47/49)
& 63.27\% (31/49)
& 32.65\% (16/49) \\

Gemma-4-31b-it (\textbf{FT})
& 89.80\% (44/49)
& 36.73\% (18/49)
&22.45\% (11/49) \\
Gemma-4-31b-it (\textbf{FT}, \textbf{Signatures})
& 93.88\% (46/49)
& 75.51\% (37/49)
& 46.94\% (23/49) \\

GPT-5
& 100.00\% (49/49)
&  44.90\% (22/49)
& 20.41\% (10/49) \\

GPT-5 (\textbf{Signatures})
& 100.00\% (49/49)
& 83.67\% (41/49)
& 51.02\% (25/49) \\

Claude-opus-4
& 100.00\% (49/49)
& 61.22\% (30/49)
& 40.82\% (20/49) \\

Claude-opus-4 (\textbf{Signatures})
& 97.96\% (48/49)
& 81.63\% (40/49)
& \textbf{57.14\% (28/49)} \\

\midrule
\multicolumn{4}{l}{\textbf{\textit{APL-to-C\# Translation with Natural Language Description (\textbf{Signatures})}}} \\
\midrule
APL $\xrightarrow{\text{Qen3-32b}}$ NL, APL+NL $\xrightarrow{\text{Qwen3-32b}}$ C\#
& 87.76\% (43/49)
& 69.39\% (34/49)
&  36.73\% (18/49) \\

APL $\xrightarrow{\text{Qwen3-32b}}$ NL, 
APL+NL $\xrightarrow{\text{GPT-5}}$ C\#
& 100.00\% (49/49)
& 83.67\% (41/49)
& 42.86\% (21/49) \\

APL $\xrightarrow{\text{GPT-5}}$ NL, 
APL+NL $\xrightarrow{\text{Qwen3-32b}}$ C\#
& 85.71\% (42/49)
& 63.27\% (31/49)
& 36.73\% (18/49) \\

APL $\xrightarrow{\text{Gemma-4-31b-it}}$ NL, 
APL+NL $\xrightarrow{\text{Gemma-4-31b-it}}$ C\#
& 89.80\% (44/49)
& 75.51\% (37/49)
& 53.06\% (26/49)  \\

APL $\xrightarrow{\text{GPT-5}}$ NL, 
APL+NL $\xrightarrow{\text{GPT-5}}$ C\#
& 100.00\% (49/49)
& 83.67\% (41/49)
& 55.10\% (27/49) \\

APL
$\xrightarrow{\text{Claude-opus-4}}$ NL, 
APL+NL $\xrightarrow{\text{Claude-opus-4}}$ C\#
& 100.00\% (48/49)
& 83.67\% (41/49)
& \textbf{59.18\% (29/49)} \\

\midrule
\multicolumn{4}{l}{\textbf{\textit{Retrieval-Augmented \textit{APL-to-C\#} Translation (Signatures)}}} \\
\midrule
GPT-5
& 97.96\% (48/49)
& 79.59\% (39/49)
& 53.06\% (26/49) \\

\midrule
\multicolumn{4}{l}{\textbf{\textit{Iterative APL-to-C\# Translation \textbf{(Signatures)}}}} \\
\midrule
Qwen3-32B (\textbf{FT})
& 97.96\% (48/49)
& 77.55\% (38/49)
& 59.18\% (29/49) \\

GPT-5
& 100.00\% (49/49)
& 87.76\% (43/49)
& 69.39\% (34/49) \\

APL
$\xrightarrow{\text{GPT-5}}$ NL, 
APL+NL $\xrightarrow{\text{GPT-5}}$ C\#
& 100.00\% (49/49)
&  87.76\% (43/49)
& 71.43\% (35/49) \\

APL
$\xrightarrow{\text{Gemma-4-31b-it}}$ NL, 
APL+NL $\xrightarrow{\text{Gemma-4-31b-it}}$ C\#
& 97.96\% (48/49)
& 87.76\% (43/49)
& 77.55\% (38/49)\\

APL$\xrightarrow{\text{Claude-opus-4}}$ NL, APL+NL
$\xrightarrow{\text{Claude-opus-4}}$ C\#
& 100.00\% (49/49)
& 91.84\% (45/49)
& \textbf{79.59\% (39/49)} \\

\bottomrule
\end{tabular}}
\end{table}

RAG provides a modest improvement for GPT-5, performing similarly to the natural language setup. Our analysis reveals that top-5 results for APL queries often share similar scores, with high normalized entropy indicating an ambiguous retrieval process. This lack of a dominant, high-utility source prevents the extraction of a meaningful context, effectively reverting the RAG mechanism into a standard natural language pipeline.

The highest performance is observed in the iterative APL-to-C\# translation setup, where both direct and natural language based iterative translations achieve the highest full pass rates, with the natural language based iterative setup with Claude-4-Opus achieves a peak full pass rate of 79.59\%, followed closely by the one with Gemma-4-31b-it at 77.55\%. This shows that automatic correction by execution feedback improves translation quality and reliability.

\begin{figure}[th!]
    \centering
    \includegraphics[width=0.9\linewidth]{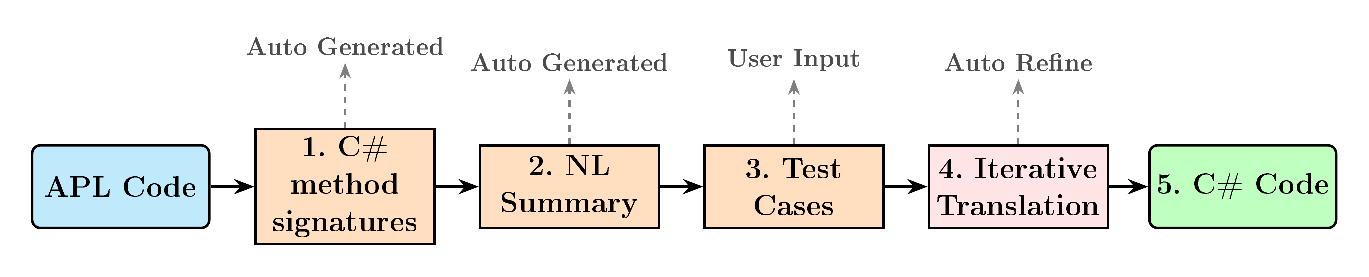}
\caption{Proposed guided APL-to-C\# translation workflow.}
\label{fig:recommended_workflow}
\end{figure}

\begin{description} 
\item[Answer to RQ1:] Overall, the results demonstrate that both model selection and setup significantly impact performance. The inclusion of method signatures consistently improve model performance. We achieve peak performance by augmenting LLMs with method signatures, NL program descriptions, and iterative feedback. Consequently, we recommend incorporating these components into APL-to-C\# translation pipelines, as illustrated in Figure~\ref{fig:recommended_workflow}.
\end{description}

\subsection{The Effect of Training Dataset Scaling} 
\label{sec:data_size_effect} 

\begin{figure}[th!]
\centering
\includegraphics[width=0.65\linewidth]{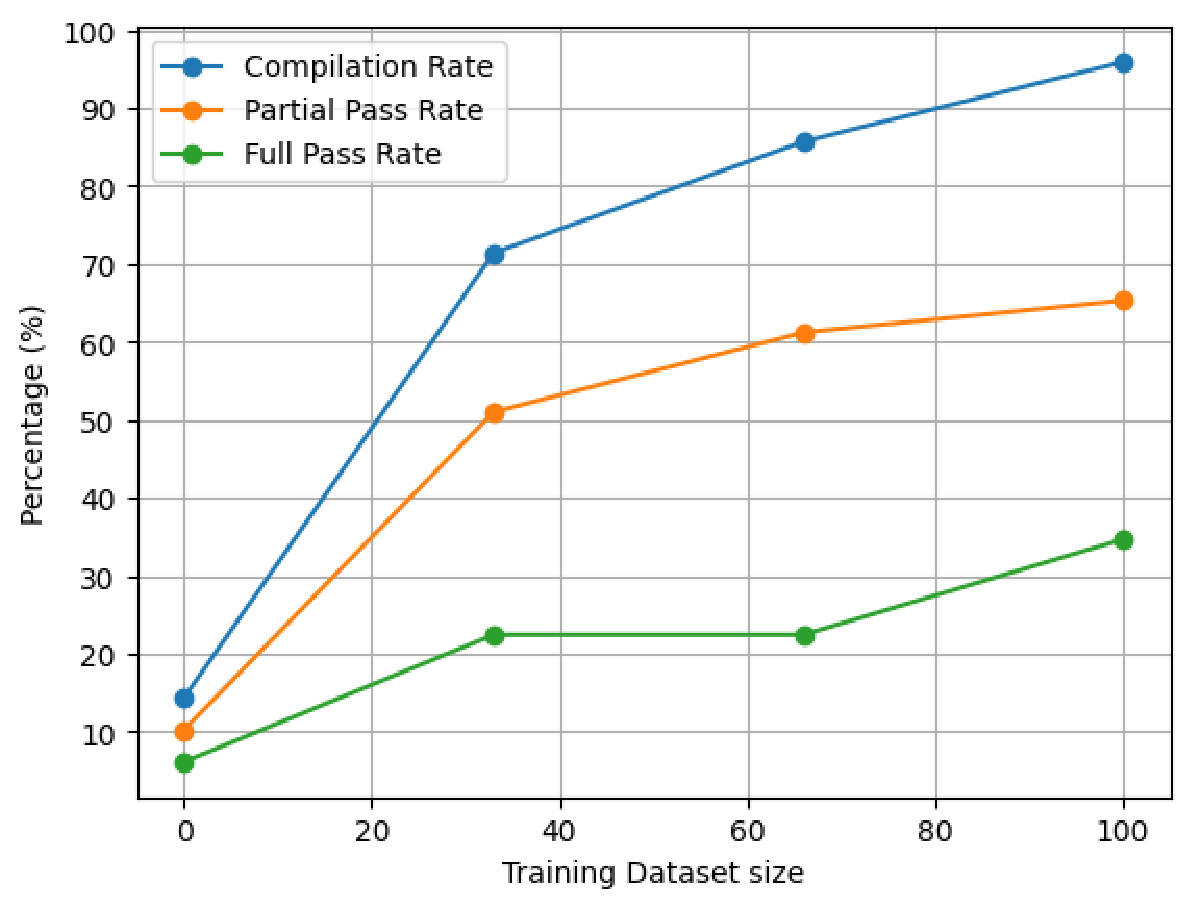}
\caption{Scaling behavior of direct APL-to-C\# translation performance relative to the size of the training set of Dataset $B$.}
\label{fig:data_scaling_results}
\end{figure}

We also evaluate how translation performance scales with the size of the training dataset. To investigate this, we conducted experiments while holding the training configuration constant and varying the size of the training set of Dataset $B$. As illustrated in Figure~\ref{fig:data_scaling_results}, increasing the dataset size leads to consistent performance gains across all evaluation metrics. Notably, while the compilation rate scales rapidly, the full pass rate shows a more gradual ascent, highlighting that complex logic requires significantly more data to master than basic syntax.

\begin{description} 
\item[Answer to RQ2:] Our findings indicate that performance is currently limited  by the dataset size; consequently, further scaling of the training dataset is likely to yield additional improvements.
\end{description}

\subsection{Classification of Translation Errors}
\label{sec:trans_err_categories}

To identify typical translation errors and understand the limitations of the evaluated models, we classify unsuccessful translations into three categories:

\begin{itemize} 
\item \textbf{Compilation errors:} The generated code does not compile due to syntax errors or invalid method signatures. 

\item \textbf{Runtime errors:} The generated code compiles successfully but fails during execution, usually due to invalid indexing or unhandled edge cases.

\item \textbf{Functional errors:} The generated code compiles and executes successfully, but produces output that deviates from the expected ground truth. 
\end{itemize}

\begin{figure}[th!]
\centering
\includegraphics[width=0.8\linewidth]{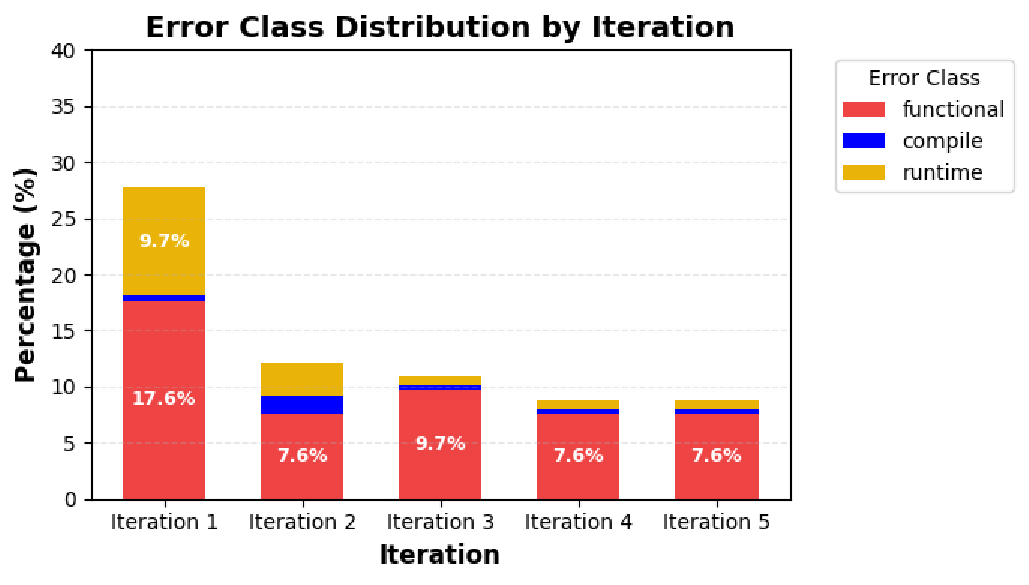}
\caption{Error class distribution across 5 iterations. Each bar shows the percentage of compilation, runtime, and functional errors, calculated across all test cases.}
\label{fig:translation-error-distribution}
\end{figure}

Figure~\ref{fig:translation-error-distribution} shows the distribution of these error categories in successive iterations. Several key observations can be made. First, functional errors constitute the dominant cause of failure, indicating that while models successfully learn to generate syntactically valid and executable code, they struggle to preserve the functionality of the original APL code. Second, compilation and runtime errors occur less frequently and are the easiest for the model to fix through iterative feedback. Finally, the success rate increases until it reaches a plateau, at which point the remaining failures are mostly related to incorrect program behavior rather than syntactic correctness.

\begin{description} 
\item[Answer to RQ3:] Overall, we observe that the majority of unsuccessful translations are caused by misunderstanding the intent of the APL code rather than low-level code generation failures.
\end{description}

\section{Conclusion}
\label{sec:conclusion}

In this work, we investigated APL-to-C\# translation using LLMs as a means of preserving legacy code functionality that exists in production systems, illustrating the inherent challenges of migrating from a dynamically typed array-oriented programming language to a statically typed object-oriented one. We implemented and evaluated four distinct translation approaches, including direct translation as a baseline, a two-stage approach that uses NL descriptions, retrieval-augmented translation, and iterative translation using compilation and execution feedback. 

Our experimental results show that while standalone pre-trained LLMs possess limited proficiency in direct APL-to-C\# translation, performance is significantly enhanced through fine-tuning combined with structural guidance, including method signatures, NL descriptions, and an iterative translation process. We have demonstrated that, even with limited training data, our optimized neural translation pipeline enables open-weight models to achieve performance competitive with closed-weight models. Consequently, this pipeline offers organizations an effective solution for modernizing legacy APL systems, thus mitigating the substantial risks and resource intensiveness inherent in manual migration efforts.

Furthermore, our pipeline can support developers in accelerating the generation of additional datasets, and a natural direction for future work is to expand the training corpus and evaluate additional LLMs. Although our findings indicate that closed-source vendor models achieve the best results (cf. Table \ref{tab:compilation_pass_rates}), transitioning to open-source alternatives is worth investigating. This would not only enable scalable, private deployments, but would also ensure greater control over intellectual property and potentially reduce computational overhead.

Another future direction is investigating the transition from a single-program translation to a multi-agent system capable of handling project-level migrations. Such a system would coordinate specialized agents to manage cross-file dependencies. This vision is supported by recent research, such as TransAgent \cite{yuan2026transagent}, LegacyTranslate \cite{Moti2026legacytranslate}, and RepoTransAgent \cite{guan2026repotrans} which demonstrate the effectiveness of multi-agent frameworks. Finally, while our focus was on APL-to-C\# translation, the neural translation approaches evaluated could also be applied to other legacy languages.

\subsubsection*{Acknowledgments.}
This work was a collaboration between  the Department of Applied Mathematics and Computer Science at the Technical University of Denmark (DTU) and SimCorp A/S, Denmark. The authors acknowledge the DTU Computing Center (DCC) for providing access to high-performance computing resources.



\begin{thebibliography}{99}
\bibitem{Vaswani2023}
Vaswani, A., Shazeer, N., Parmar, N., Uszkoreit, J., Jones, L., et al.: Attention is all you need. arXiv preprint \href{https://arxiv.org/abs/1706.03762}{arXiv:1706.03762} (2023)

\bibitem{Wu2016}
Wu, Y., Schuster, M., Chen, Z., Le, Q.V., Norouzi, M., et al.: Google's neural machine translation system: bridging the gap between human and machine translation. arXiv preprint \href{https://arxiv.org/abs/1609.08144}{arXiv:1609.08144} (2016)

\bibitem{DellAcqua2023}
Dell'Acqua, F., McFowland, E., III, Mollick, E., Lifshitz-Assaf, H., Kellogg, K.C., et al.: Navigating the jagged technological frontier: Field experimental evidence of the effects of AI on knowledge worker productivity and quality. Harvard Business School, Working Paper 24-013 (2023)

\bibitem{Iverson1962}
Iverson, K.E.: A programming language. Wiley (1962)

\bibitem{Lei2023}
Lei, B., Ding, C., Chen, L., Lin, P.-H., Liao, C.: Creating a dataset for high-performance computing code translation using LLMs: a bridge between OpenMP Fortran and C++. arXiv preprint \href{https://arxiv.org/abs/2307.07686}{arXiv:2307.07686} (2023)

\bibitem{Hu2022}
Hu, E.J., Shen, Y., Wallis, P., Allen-Zhu, Z., Li, Y., et al.: LoRA: low-rank adaptation of large language models. In: International Conference on Learning Representations (ICLR) (2022)

\bibitem{Ruiz2025}
Vallecillos Ruiz, F., Hort, M., Moonen, L.: The art of repair: optimizing iterative program repair with instruction-tuned models. arXiv preprint \href{https://arxiv.org/abs/2505.02931}{arXiv:2505.02931} (2025)

\bibitem{pan2023lost}
R.~Pan, A.~R.~Ibrahimzada, R.~Krishna, D.~Sankar, L.~P.~Wassi, M.~Merler, B.~Sobolev, R.~Pavuluri, S.~Sinha, and R.~Jabbarvand,
``Lost in Translation: A Study of Bugs Introduced by Large Language Models while Translating Code,''
\emph{arXiv preprint arXiv:2308.03109}, 2023.

\bibitem{Aljagthami2025}
Aljagthami, A., Banabila, M., Alshehri, M., Kabini, M., Alahmadi, M.D.: Evaluating large language models for code translation: effects of prompt language and prompt design. arXiv preprint \href{https://arxiv.org/abs/2509.12973}{arXiv:2509.12973} (2025)

\bibitem{Chen2025}
Chen, X., Xue, J., Xie, X., Liang, C., Ju, X.: A systematic literature review on neural code translation. arXiv preprint \href{https://arxiv.org/abs/2505.07425}{arXiv:2505.07425} (2025)

\bibitem{Lu2021}
Lu, S., Guo, D., Ren, S., Huang, J., Svyatkovskiy, A., et al.: CodeXGLUE: a machine learning benchmark dataset for code understanding and generation. arXiv preprint \href{https://arxiv.org/abs/2102.04664}{arXiv:2102.04664} (2021)

\bibitem{Puri2021}
Puri, R., Kung, D.S., Janssen, G., Zhang, W., Domeniconi, G., et al.: CodeNet: a large-scale AI for code dataset for learning a diversity of coding tasks. arXiv preprint \href{https://arxiv.org/abs/2105.12655}{arXiv:2105.12655} (2021)

\bibitem{Ahmad2023Avatar}
Ahmad, W.U., Tushar, M.G.R., Chakraborty, S., Chang, K.-W.: AVATAR: a parallel corpus for Java-Python program translation. arXiv preprint \href{https://arxiv.org/abs/2108.11590}{arXiv:2108.11590} (2023)

\bibitem{Chen2018}
Chen, X., Liu, C., Song, D.: Tree-to-tree neural networks for program translation. arXiv preprint \href{https://arxiv.org/abs/1802.03691}{arXiv:1802.03691} (2018)

\bibitem{IBM2023}
IBM: IBM unveils Watsonx generative AI capabilities to accelerate mainframe application modernization. IBM Newsroom (2023)

\bibitem{Ahmad2023Backtranslate}
Ahmad, W.U., Chakraborty, S., Ray, B., Chang, K.-W.:
Summarize and Generate to Back-translate: Unsupervised Translation of Programming Languages.
arXiv preprint \href{https://arxiv.org/abs/2205.11116}{arXiv:2205.11116} (2023)

\bibitem{Tai2025NL}
Tai, C.A., Nie, P., Golab, L., Wong, A.: NL in the middle: code translation with LLMs and intermediate representations.
arXiv preprint \href{https://arxiv.org/abs/2507.08627}{arXiv:2507.08627} (2025)

\bibitem{Bhattarai2024}
Bhattarai, M., Santos, J.E., Jones, S., Biswas, A., Alexandrov, B., O'Malley, D.: Enhancing code translation in language models with few-shot learning via retrieval-augmented generation.
arXiv preprint \href{https://arxiv.org/abs/2407.19619}{arXiv:2407.19619} (2024)

\bibitem{Ibrahimzada2025}
Ibrahimzada, A.R., Ke, K., Pawagi, M., Abid, M.S., Pan, R., et al.: AlphaTrans: A neuro-symbolic compositional approach for repository-level code translation and validation. Proceedings of the ACM on Software Engineering 2(FSE), 2454--2476 (2025)

\bibitem{dyalog_oo_apl}
Dyalog Ltd.: Object oriented programming for APL programmers. (2017) \url{https://docs.dyalog.com/19.0/Object%20Oriented%20Programming%20for%20APL%20Programmers.pdf}, last accessed 2026/04/12

\bibitem{Lewis2021}
Lewis, P., Perez, E., Piktus, A., Petroni, F., Karpukhin, V., et al.: Retrieval-augmented generation for knowledge-intensive NLP tasks. arXiv preprint \href{https://arxiv.org/abs/2005.11401}{arXiv:2005.11401} (2021)

\bibitem{rosetta_code}
Rosetta Code: Programming Tasks. \url{https://rosettacode.org/wiki/Category:Programming_Tasks}, last accessed 2026/04/12

\bibitem{huggingface}
Hugging Face, Inc.: Hugging Face: The AI community building the future.
\url{https://huggingface.co}, last accessed 2026/04/09 

\bibitem{gemma4}
Google DeepMind. \textit{Gemma 4}. 2026. Available at: \url{https://deepmind.google/models/gemma/gemma-4/}

\bibitem{yang2025qwen3technicalreport}
Yang, A., Li, A., Yang, B., Zhang, B., Hui, B., et al.: Qwen3 technical report. arXiv preprint \href{https://arxiv.org/abs/2505.09388}{arXiv:2505.09388} (2025)

\bibitem{gpt5_chat}
OpenAI: GPT-5-chat-latest documentation. \url{https://platform.openai.com/docs/models/gpt-5-chat-latest}, last accessed 2026/04/12

\bibitem{anthropic_claude4}
Anthropic: Claude 4 Opus documentation. \url{https://www.anthropic.com/news/claude-4-opus}, last accessed 2026/04/12

\bibitem{apl2_idioms}
Cason, S.: APL2 idioms library. Technical Report, IBM Corporation (1989)

\bibitem{finnapl}
APL Wiki. FinnAPL idiom library. Finnish APL Association.
\url{https://aplwiki.com/wiki/FinnAPL_idiom_library}, last accessed 2026/04/12 

\bibitem{openai_embedding}
OpenAI: Text-embedding-3-large documentation. \url{https://platform.openai.com/docs/models/text-embedding-3-large}, last accessed 2026/04/12

\bibitem{gpt5_mini}
OpenAI: GPT-5-mini documentation. \url{https://platform.openai.com/docs/models/gpt-5-mini}, last accessed 2026/04/12

\bibitem{yuan2026transagent}
Yuan, Z., Chen, W., Wang, H., Peng, X., Chen, Z., Lou, Y.:
TransAgent: Enhancing LLM-based code translation via fine-grained execution alignment. arXiv preprint \href{https://arxiv.org/abs/2409.19894v5}{arXiv:2409.19894v5} (2026)

\bibitem{Moti2026legacytranslate}
Moti, Z., Soudani, H., van der Kogel, J.: LegacyTranslate: LLM-based multi-agent method for legacy code translation. arXiv preprint \href{https://arxiv.org/abs/2603.14054}{arXiv:2603.14054} (2026)

\bibitem{guan2026repotrans}
Guan, Z., Yin, X., Peng, Z., Ni, C.: RepoTransAgent: Multi-agent LLM framework for repository-aware code translation. In: Proceedings of the 48th International Conference on Software Engineering (2026)

\end{thebibliography}
\end{document}